\documentclass[12pt,a4 paper]{article}
\usepackage{amssymb}
\usepackage{amsmath}
\usepackage[dvips]{graphicx}
\begin{document}
\noindent \textbf{\Large  Bianchi type VI cosmological models: A Scale-Covariant study}\\

\vspace{0.5cm}
\noindent ${{\textbf{Mohd. Zeyauddin}^{}}}$ \footnote{ corresponding author}.${{\textbf {Bijan Saha}}^{2}}$  \\

\noindent $^{1}$ Bogoliubov Laboratory of Theoretical Physics,\\
Joint Institute For Nuclear Research, Dubna - 141980,\\
Moscow Region Russia\\
\noindent E-mail: $^2$ zeya@theor.jinr.ru,

\vspace{0.5cm}

\noindent $^{2}$ Laboratory of Information Technologies,\\
Joint Institute For Nuclear Research, Dubna - 141980,\\
Moscow Region Russia\\
\noindent E-mail: $^1$ bijan@jinr.ru,\\

\vspace {0.5cm} \noindent \textbf{Abstract} A model for an
anisotropic Bianchi type VI universe in a Scale Covariant theory of
gravitation (Canuto et al. 1977) is analyzed. Exact solutions to the
corresponding field equations are found under some specific
assumptions. A finite singularity is found in the model at the
initial time $t=0$. All the physical parameters are studied and
thoroughly discussed. The model behaves like a big bang singular model o
f the universe.  \\


\pagebreak
\pagestyle{headings}
\noindent \textbf{1 Introduction}\\

\noindent Canuto et al. (1977) have formulated a Scale-Covariant
theory of gravitation by associating the mathematical operation of
scale transformation with the physics of using different dynamical
systems to measure space-time distances. A Scale-Covariant theory
provides the necessary theoretical framework to sensibly discuss the
possible variation of the gravitational constant $G$ without
compromising the validity of general relativity. In this theory, we
measure physical quantities in atomic units whereas Einstein's field
equations in gravitational units. If we consider
$d\bar{s}^{2}=\bar{g}_{ij}dx^{i}dx^{j}$, the line element in
Einstein units, the corresponding line element in any other units
(in atomic units) will be written as

\begin{equation}
ds=\phi^{-1}(x)d\bar{s}.
\end{equation}
The metric tensor in the two systems of units are related by a conformal transformation

\begin{equation}
\bar{g}_{ij}=\phi^{2}g_{ij},
\end{equation}
where the metric $\bar{g}_{ij}$ giving macroscopic metric properties and $g_{ij}$ giving
microscopic metric properties. Here we consider the gauge function $\phi$ as a function of time.\\

\noindent Friedmann-Robertson Walker(FRW) space-time models are
widely acceptable as a good approximation of the present stage of
the evolution of the universe although it is spatially homogeneous
and isotropic in nature. However, the large scale matter
distribution in the observable universe, largely manifested in the
form of discrete structures, does not exhibit a high degree of
homogeneity. Also the recent space investigations detect anisotropy
in the cosmic microwave background. So the recent experimental data
support the existence of an anisotropic phase that approaches an
isotropic phase. These theoretical arguments (Saha 2004) lead one to
consider models with an anisotropic background. Bianchi type
space-times play a vital role in understanding and description of
the early stages of evolution of the universe.
Bianchi type VI (Saha 2004) space-time is inhomogeneous and anisotropic. \\

\noindent Scale-Covariant theory in different Bianchi space-times has been studied so far by
several authors. Shri Ram et al. (2009) have studied a spatially homogeneous Bianchi type V
cosmological model in Scale-Covariant theory of gravitation. Reddy et al. (2007) have developed
a cosmological model with negative constant deceleration parameter in Scale-Covariant theory of
gravitation. Beesham (1986) has obtained a solution for Bianchi type I cosmological model in the
Scale-Covariant theory. Higher dimensional string cosmologies in Scale-Covariant theory of gravitation
have been investigated by Venkateswarlu and Kumar (2004). Reddy et al. (1993) have presented the exact
Bianchi type II, VIII and IX cosmological models in Scale-Covariant theory of gravitation. In this paper,
we obtain exact solution to the field equations of Scale-Covariant theory for Bianchi type VI space-time metric.\\\\

\noindent \textbf{2  Field Equations, Metric and General Expressions}\\

\noindent Canuto et al. (1977) transformed the general Einstein's field equations by using the conformal
transformations equations (1) and (2) as follows:

\begin{equation}
R_{\mu\nu}-\frac{1}{2}g _{\mu\nu}R+f_{\mu\nu}(\phi) =-8\pi G T_{\mu\nu}+\Lambda (\phi)g_{\mu\nu},
\end{equation}

where

\begin{equation}
\phi^{2}f_{\mu\nu}= 2\phi \phi_{\mu;\nu}-4\phi_{\mu}\phi_{\nu}-g_{\mu\nu}(\phi\phi^{\lambda}_{;\lambda}-\phi^{\lambda}\phi_{\lambda}),
\end{equation}
for any scalar $\phi, \indent  \phi_{\mu}=\phi_{,\mu}$. Here comma denotes ordinary partial differentiation whereas a
semi-colon denotes a covariant differentiation. \\

\noindent The Bianchi VI space-time metric is given as

\begin{equation}
ds^2=dt^2-A^{2} dx^2-e^{-2m x}B^{2}dy^2 - e^{2n x}C^{2}dz^2,
\end{equation}
with the scale factors $A$, $B$, $C$ being functions of time only.
Here $m$ ,$n$ are some arbitrary constants. Here the source of
gravitational field is considered as a perfect fluid. So for a
perfect fluid, the energy momentum tensor is given by

\begin{equation}
T_{\mu\nu} = \left(\rho +p \right)u_{\mu}u_{\nu} - p g_{\mu\nu},
\end{equation}
where $\rho$ is the energy-density, $p$ the pressure and $u^{\mu}$ is the four velocity vector of
the fluid following $u^{\mu}u_{\mu}=1$.\\

\noindent The general formulas of certain physical parameters for the metric equation (5) are given as follows:\\

\noindent The expansion scalar is given by

\begin{equation}
\theta = u ^{\mu}_{;\mu}=\frac{\dot{A}}{A}+\frac{\dot{B}}{B}+\frac{\dot{C}}{C},
\end{equation}
where a dot$(.)$ denotes differentiation with respect to time $t$. The shear scalar has the form

\begin{equation}
\sigma ^{2} = \frac {1}{2}\sigma _{\mu\nu}\sigma^{\mu\nu}= \frac {1}{2}\left[\left(\frac{\dot{A}}{A}\right)^{2}+  \left(\frac{\dot{B}}{B}\right)^{2}+\left(\frac{\dot{C}}{C}\right)^{2}\right]-\frac{{\theta}^{2}}{6}.
\end{equation}
We also introduce generalized Hubble parameter $H$:
\begin{equation}
H= \frac{1}{3}\left(H_{1}+H_{2}+H_{3}\right),
\end{equation}
with $H_{1}=\frac{\dot{A}}{A}$, $H_{2}=\frac{\dot{B}}{B}$ and
$H_{3}=\frac{\dot{C}}{C}$  are the directional Hubble parameters  in the
directions of $x$, $y$ and $z$ respectively. Let us introduce the
function $V$ and average scale factor $a$:
\begin{equation}
V=ABC,
\end{equation}

\begin{equation}
a=(ABC)^{1/3}.
\end{equation}
It should be noted that the parameters $H$, $V$ and $a$ are connected by the following relation

\begin{equation}
H=\frac{1}{3}\frac{\dot{V}}{V}=\frac{\dot{a}}{a}.
\end{equation}
The field equations (3) and (4) to the metric equation (5) for perfect fluid equation (6), are given as following set of equations

\begin{eqnarray}
\frac{\ddot{B}}{B}+\frac{\ddot{C}}{C}+\frac{\dot{B}}{B}\frac{\dot{C}}{C}+\frac{mn}{A^{2}}-2\frac{\dot{A}}{A}\frac{\dot{\phi}}{\phi}
+\frac{\dot{\phi}}{\phi}\left(\frac{\dot{V}}{V}\right)+\frac{\ddot{\phi}}{\phi}-\frac{\dot{\phi}^{2}}{\phi^{2}}=-8\pi G p,
\end{eqnarray}

\begin{eqnarray}
\frac{\ddot{A}}{A}+\frac{\ddot{C}}{C}+\frac{\dot{A}}{A}\frac{\dot{C}}{C}-\frac{n^2}{A^{2}}-2\frac{\dot{B}}{B}\frac{\dot{\phi}}{\phi}
+\frac{\dot{\phi}}{\phi}\left(\frac{\dot{V}}{V}\right)+\frac{\ddot{\phi}}{\phi}-\frac{\dot{\phi}^{2}}{\phi^{2}}=-8\pi G p,
\end{eqnarray}

\begin{eqnarray}
\frac{\ddot{A}}{A}+\frac{\ddot{B}}{B}+\frac{\dot{A}}{A}\frac{\dot{B}}{B}-\frac{m^{2}}{A^{2}}-2\frac{\dot{C}}{C}\frac{\dot{\phi}}{\phi}
+\frac{\dot{\phi}}{\phi}\left(\frac{\dot{V}}{V}\right)+\frac{\ddot{\phi}}{\phi}-\frac{\dot{\phi}^{2}}{\phi^{2}}=-8\pi G p,
\end{eqnarray}

\begin{eqnarray}
\frac{\dot{A}}{A}\frac{\dot{B}}{B}+\frac{\dot{A}}{A}\frac{\dot{C}}{C}+\frac{\dot{B}}{B}\frac{\dot{C}}{C}-\frac{m^{2}-mn+n^{2}}{A^{2}}
+\frac{\dot{\phi}}{\phi}\left(\frac{\dot{V}}{V}\right)-\frac{\ddot{\phi}}{\phi}+3\frac{\dot{\phi}^{2}}{\phi^{2}}= 8\pi G \rho,
\end{eqnarray}

\begin{eqnarray}
m\frac{\dot{B}}{B}-n\frac{\dot{C}}{C}-(m-n)\frac{\dot{A}}{A}=0.
\end{eqnarray}
Here we have used definition (10). The Bianchi identity reads

\begin{eqnarray}
\dot{\rho}+(\rho+p) \frac{\dot{V}}{V}+
\rho\frac{\dot{\phi}}{\phi}+3p\frac{\dot{\phi}}{\phi}=0.
\end{eqnarray}
From equation (17), we find the following relation between the metric functions $A$, $B$, $C$ as

\begin{eqnarray}
\left(\frac{B}{A}\right)^{m}=k\left(\frac{C}{A}\right)^{n},
\end{eqnarray}
with the integration constant $k$. Taking into account the definition (10), from equation (19), we can write the scale factors $B$ and $C$ in terms of $A$ and $V$, such that

\begin{eqnarray}
B=\left(k V^{n}A^{(m-2n)}\right)^\frac{1}{m+n},
\end{eqnarray}

\begin{eqnarray}
C=\left(\frac{1}{k} V^{m}A^{(n-2m)}\right)^\frac{1}{m+n}.
\end{eqnarray}
Summing equations (13), (14), (15) and 3 times equation (16), in view of the equation (10) for volume scalar, we obtain a non-linear differential equation as

\begin{eqnarray}
\frac{\ddot{V}}{V}+\frac{2}{A^{2}}(mn-n^{2}-m^{2})+2\frac{\dot{\phi}}{\phi}\frac{\dot{V}}{V}+3\frac{\dot{\phi}^{2}}{\phi^{2}}
=12\pi G (-p+\rho).
\end{eqnarray}
Taking into account that the perfect fluid obeys the equation of state $p=\gamma \rho, (0<\gamma<1)$, the equation (18) becomes

\begin{eqnarray}
\rho V^{1+\gamma}\phi^{1+3\gamma}=a_{0}
\end{eqnarray}
where $a_0$ is an integration constant. \\

\noindent We consider the gauge function $\phi$ (Canuto et al. (1977) and Shri Ram et al. (2009)) as

\begin{eqnarray}
\phi=\phi_{0}a^{\alpha}=\phi_{0}V^{\alpha/3},
\end{eqnarray}
where $\alpha$ and $\phi_{0}$ are arbitrary constants.  Now in view of equations (23), (24) equation (22) reduces to

\begin{eqnarray}
\frac{\ddot{V}}{V}+ \frac{\alpha(\alpha+2)}{3}\left(\frac{\dot{V}}{V}\right)^2+\frac{2}{A^{2}}(mn-n^{2}-m^{2})
= \frac{12\pi G (1-\gamma)\rho_{0}}{V^{1+\gamma+\alpha/3+\alpha\gamma}}
\end{eqnarray}
where $\rho_{0}$ is an arbitrary constant. As we can see, there are
two unknown functions $A$ and $V$ in the above equation (25). Let us
demand an additional assumption relating to these two variables. So
we consider here that the scale factor $A$ is related to the volume
scalar $V$ with the relation $A=\sqrt V$ (Saha 2004). This
assumption provide us the exact solutions to the field equations at
the
same time leaving the spacetime anisotropic.\\

\noindent Note that such an assumption imposes restrictions on the metric functions.
Now, in what follows, we try to find an exact solution of the field equations in
Scale-Covariant theory with the help of the equation (25).\\

\noindent \textbf{3  Exact Solutions}\\

\noindent Under the assumption $A=\sqrt V$, we obtain the following equation for $V$, by solving the differential equation (25) as

\begin{eqnarray}
V=\frac{3v_{0}}{2(\alpha^2+2\alpha+3)}t^2+\frac{3v_1}{3-\alpha^2-2\alpha}t^{\frac{3-\alpha^2-2\alpha}{3}}+v_2,
\end{eqnarray}
where $v_0=12\pi G(1-\gamma)\rho_0+2(m^2+n^2-mn) >0$, $v_1$ and
$v_2$ are integration constants. It should be noted that in case of
a non-zero $v_2$, $V$ is non-trivial even at $t = 0$, which imposes
that $v_2$ is essentially positive. For $v_2 = 0$ we have the model,
when $V$ becomes zero at the initial time, i.e., $V\bigl|_{v_2 = 0,
t= 0} = 0$.
We also have a relationship between $\gamma$ and $\alpha$ as $\alpha=-\frac{3\gamma}{3\gamma + 1}$,  $\alpha \in\left(-3/4,0\right)$.\\

\noindent The equation (26), in view of (10), gives the following
expressions of the scale factors $A$, $B$ and $C$ as follows:

\begin{eqnarray}
A=\left[\frac{3v_{0}}{2(\alpha^2+2\alpha+3)}t^2+\frac{3v_1}{3-\alpha^2-2\alpha}t^{\frac{3-\alpha^2-2\alpha}{3}}+v_2\right]^{1/2},
\end{eqnarray}

\begin{eqnarray}
B=B_0\left[\frac{3v_{0}}{2(\alpha^2+2\alpha+3)}t^2+\frac{3v_1}{3-\alpha^2-2\alpha}t^{\frac{3-\alpha^2-2\alpha}{3}}+v_2\right]^{\frac{m}{2(m+n)}},
\end{eqnarray}

and

\begin{eqnarray}
C=C_0\left[\frac{3v_{0}}{2(\alpha^2+2\alpha+3)}t^2+\frac{3v_1}{3-\alpha^2-2\alpha}t^{\frac{3-\alpha^2-2\alpha}{3}}+v_2\right]^{\frac{n}{2(m+n)}},
\end{eqnarray}
where  $B_0=k^{1/(m+n)}$ and $C_0=\left(1/k\right)^{1/(m+n)}$. \\

\noindent The expressions for the gauge function $\phi$ and the average scale factor $a$ are given by

\begin{eqnarray}
\phi=\phi_0\left[\frac{3v_{0}}{2(\alpha^2+2\alpha+3)}t^2+\frac{3v_1}{3-\alpha^2-2\alpha}t^{\frac{3-\alpha^2-2\alpha}{3}}+v_2\right]^{\alpha/3},
\end{eqnarray}

and

\begin{eqnarray}
a=\left[\frac{3v_{0}}{2(\alpha^2+2\alpha+3)}t^2+\frac{3v_1}{3-\alpha^2-2\alpha}t^{\frac{3-\alpha^2-2\alpha}{3}}+v_2\right]^{1/3}.
\end{eqnarray}
Using the above expressions in equations (7)-(9), the expansion scalar $\theta$, shear scalar $\sigma^2$ and the Hubble parameter $H$ are written as,

\begin{eqnarray}
\theta=\frac{\frac{3v_0}{(\alpha^2+2\alpha+3)}t^{\frac{3+\alpha^2+2\alpha}{3}}+v_1}{\frac{3v_{0}}{2(\alpha^2+2\alpha+3)}t^{\frac{6+\alpha^2+2\alpha}{3}}+\frac{3v_1}{3-\alpha^2-2\alpha}t+v_2t^{\frac{\alpha^2+2\alpha}{3}}},
\end{eqnarray}

\begin{eqnarray}
\sigma^2=\frac{(m^2+n^2-mn)}{12(m+n)^2}\left[\frac{\frac{3v_0}{(\alpha^2+2\alpha+3)}t^{\frac{3+\alpha^2+2\alpha}{3}}+v_1}{\frac{3v_{0}}{2(\alpha^2+2\alpha+3)}t^{\frac{6+\alpha^2+2\alpha}{3}}+\frac{3v_1}{3-\alpha^2-2\alpha}t+v_2t^{\frac{\alpha^2+2\alpha}{3}}}\right]^2,
\end{eqnarray}

and

\begin{eqnarray}
H=\frac{1}{3}\left[\frac{\frac{3v_0}{(\alpha^2+2\alpha+3)}t^{\frac{3+\alpha^2+2\alpha}{3}}+v_1}{\frac{3v_{0}}{2(\alpha^2+2\alpha+3)}t^{\frac{6+\alpha^2+2\alpha}{3}}+\frac{3v_1}{3-\alpha^2-2\alpha}t+v_2t^{\frac{\alpha^2+2\alpha}{3}}}\right].
\end{eqnarray}
The directional Hubble parameters can be obtained as

\begin{eqnarray}
H_1=\frac{\frac{3v_0}{(\alpha^2+2\alpha+3)}t^{\frac{3+\alpha^2+2\alpha}{3}}+v_1}{\frac{3v_{0}}{(\alpha^2+2\alpha+3)}t^{\frac{6+\alpha^2+2\alpha}{3}}+\frac{6v_1}{3-\alpha^2-2\alpha}t+2v_2t^{\frac{\alpha^2+2\alpha}{3}}},
\end{eqnarray}

\begin{eqnarray}
H_2=\frac{\frac{3mv_0}{(m+n)(\alpha^2+2\alpha+3)}t^{\frac{3+\alpha^2+2\alpha}{3}}+\frac{m v_1}{(m+n)}}{\frac{3v_{0}}{(\alpha^2+2\alpha+3)}t^{\frac{6+\alpha^2+2\alpha}{3}}+\frac{6v_1}{3-\alpha^2-2\alpha}t+2v_2t^{\frac{\alpha^2+2\alpha}{3}}},
\end{eqnarray}

and

\begin{eqnarray}
H_3=\frac{\frac{3nv_0}{(m+n)(\alpha^2+2\alpha+3)}t^{\frac{3+\alpha^2+2\alpha}{3}}+\frac{n v_1}{(m+n)}}{\frac{3v_{0}}{(\alpha^2+2\alpha+3)}t^{\frac{6+\alpha^2+2\alpha}{3}}+\frac{6v_1}{3-\alpha^2-2\alpha}t+2v_2t^{\frac{\alpha^2+2\alpha}{3}}}.
\end{eqnarray}
Now the value of the energy-momentum tensor $\rho$ and the pressure $p$ can be found as follows:

\begin{eqnarray}
\rho= \frac{\rho_0}{\frac{3v_{0}}{2(\alpha^2+2\alpha+3)}t^2+\frac{3v_1}{3-\alpha^2-2\alpha}t^{\frac{3-\alpha^2-2\alpha}{3}}+v_2},
\end{eqnarray}

and

\begin{eqnarray}
p= \frac{\gamma\rho_0}{\frac{3v_{0}}{2(\alpha^2+2\alpha+3)}t^2+\frac{3v_1}{3-\alpha^2-2\alpha}t^{\frac{3-\alpha^2-2\alpha}{3}}+v_2}.
\end{eqnarray}

\begin{figure}[ht]
\begin{minipage}{0.5\linewidth}
\hspace{-1.cm}
\includegraphics[width=1.15\textwidth]{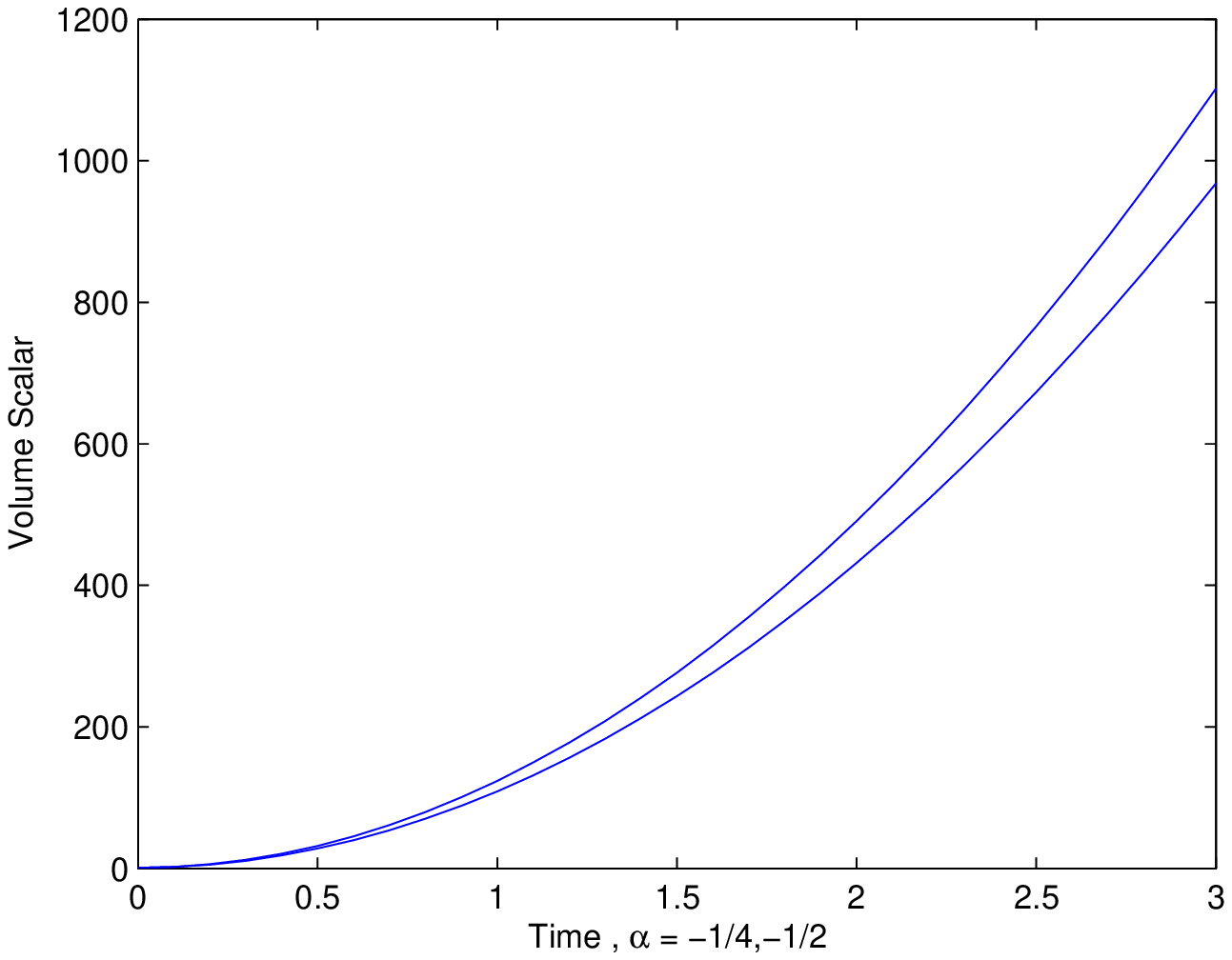}
\caption{Variation of Volume scalar $V$ with time $t$.}
\end{minipage}
\hspace{0.25cm}
\begin{minipage}{0.5\linewidth}
\includegraphics[width=1.15\textwidth]{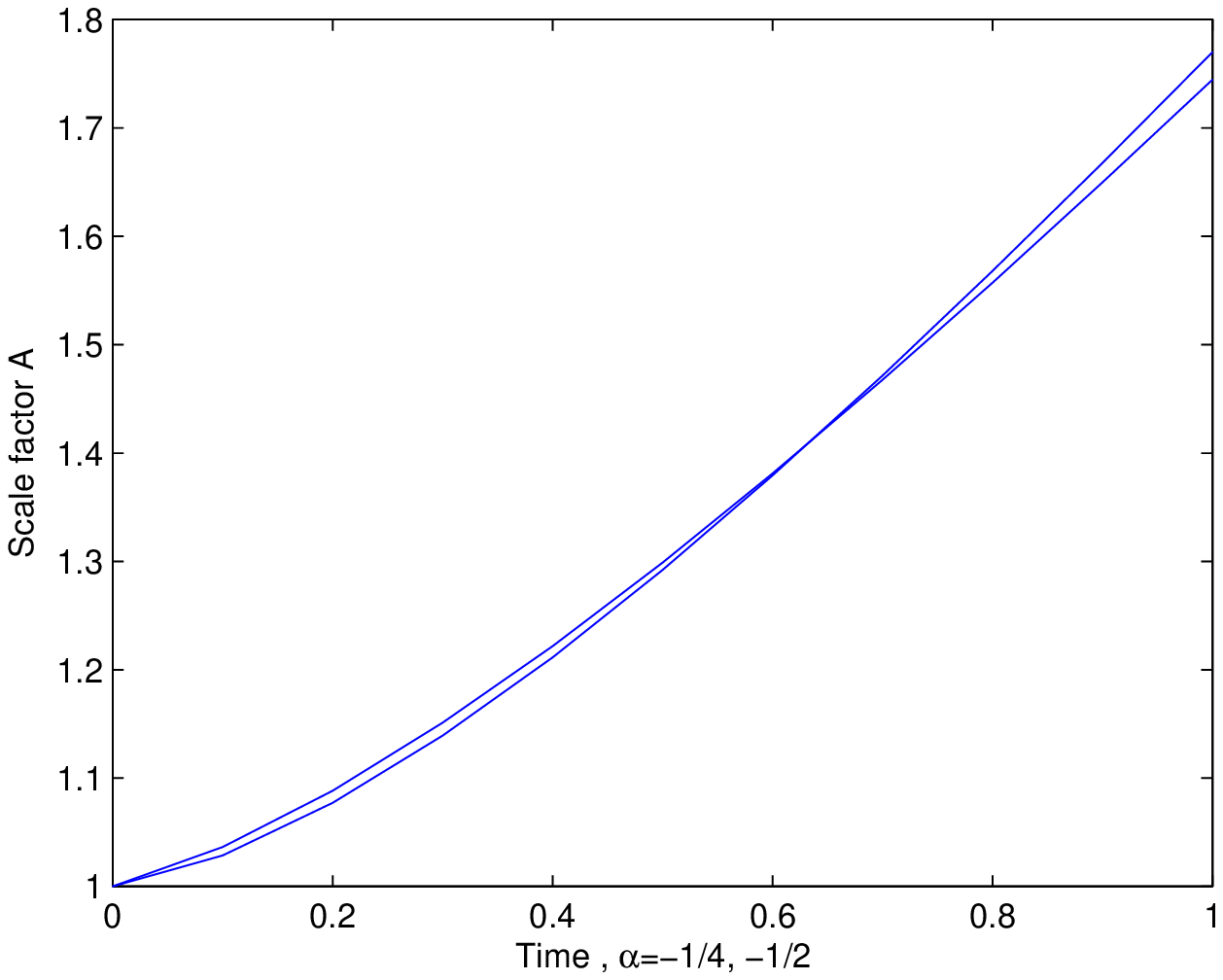}
\caption{Variation of the Scale factor $A$ with time $t$.}
\end{minipage}
\label{fig:1}
\end{figure}

\begin{figure}[ht]
\begin{minipage}{0.5\linewidth}
\hspace{-1.cm}
\includegraphics[width=1\textwidth]{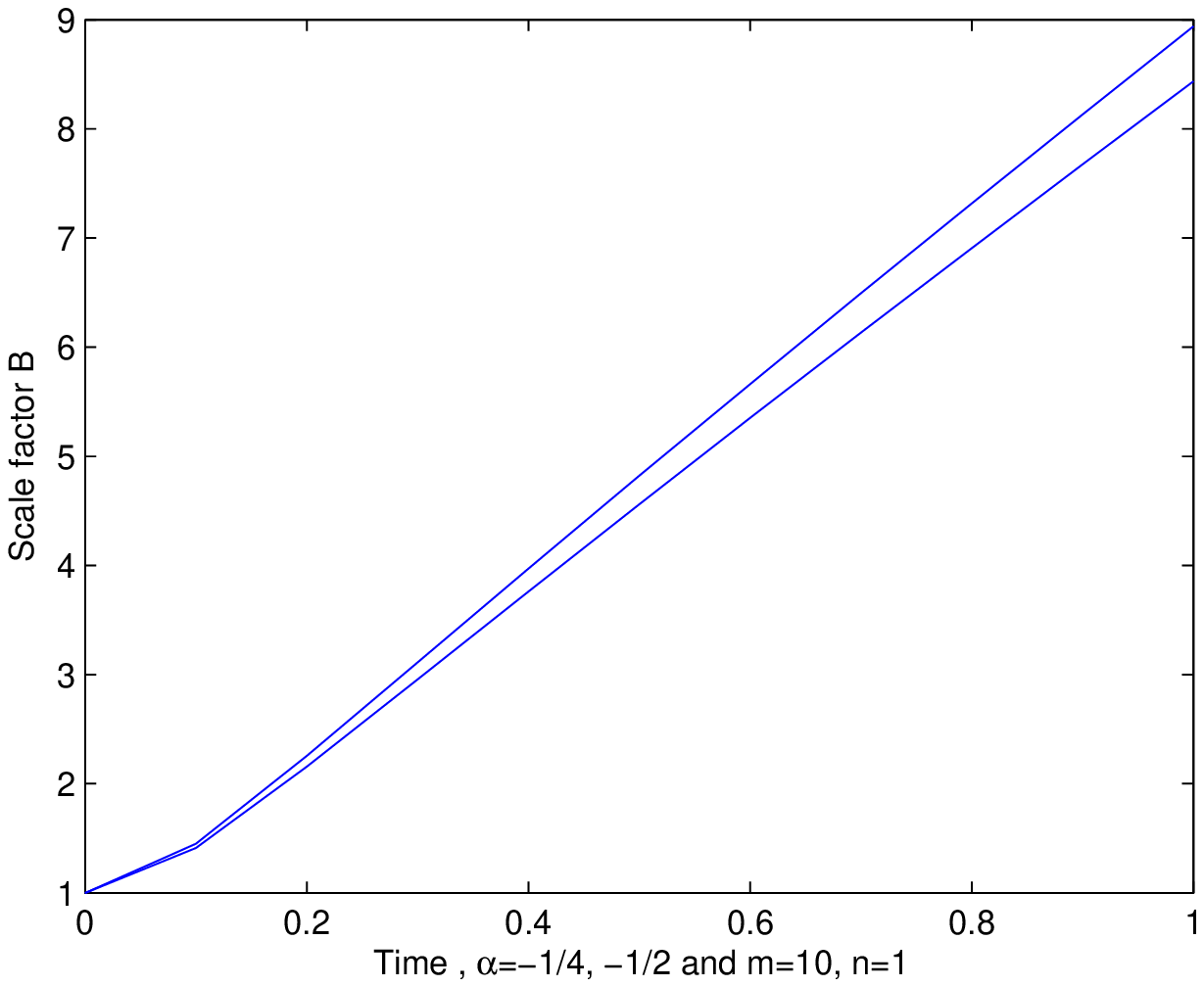}
\caption{Variation of the Scale factor $B$ with time $t$ .}
\end{minipage}
\hspace{0.25cm}
\begin{minipage}{0.5\linewidth}
\includegraphics[width=1\textwidth]{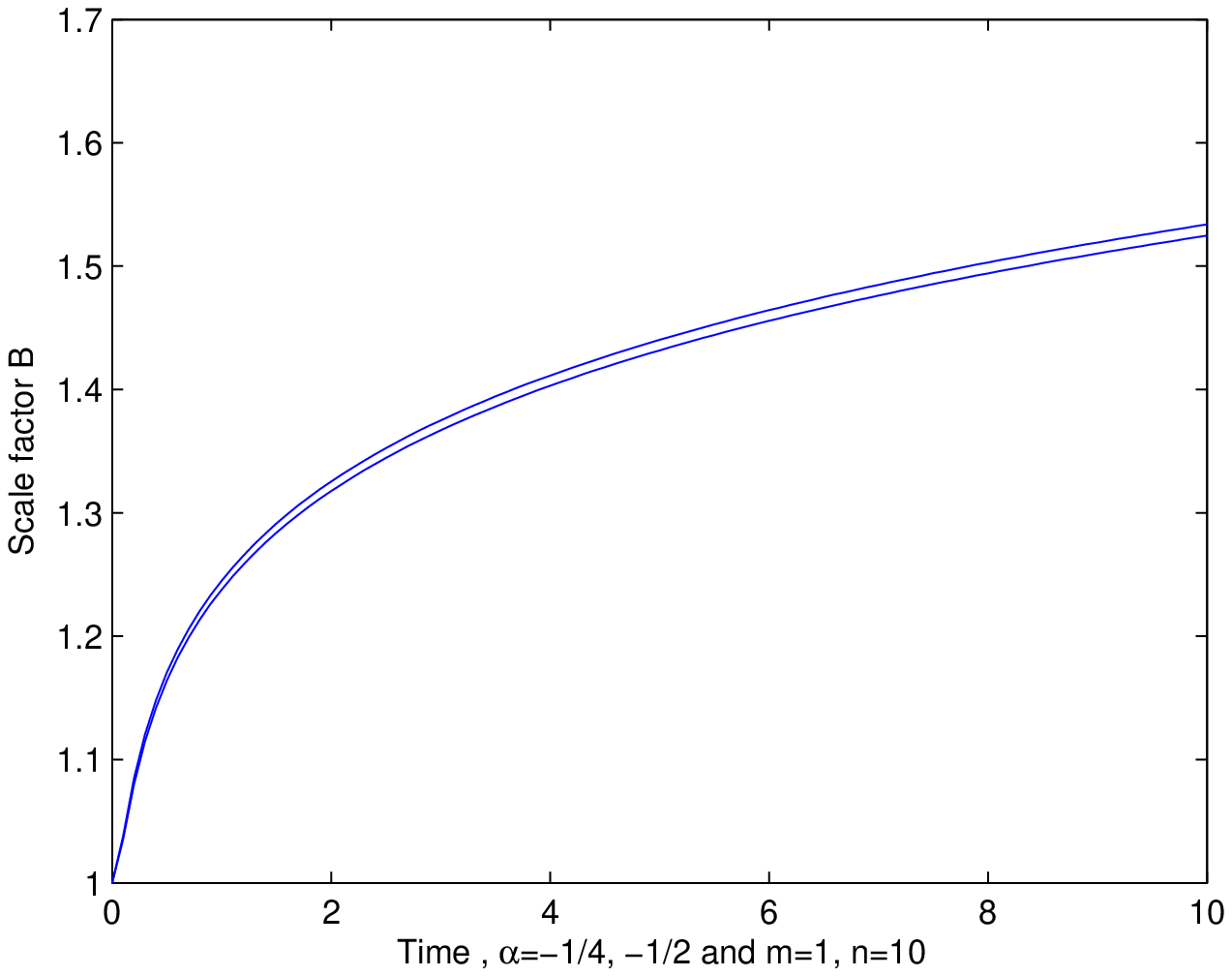}
\caption{Variation of the Scale factor $B$ with time $t$.}
\end{minipage}
\label{fig:2}
\end{figure}

\begin{figure}[ht]
\begin{minipage}{0.5\linewidth}
\hspace{-1.cm}
\includegraphics[width=0.95\textwidth]{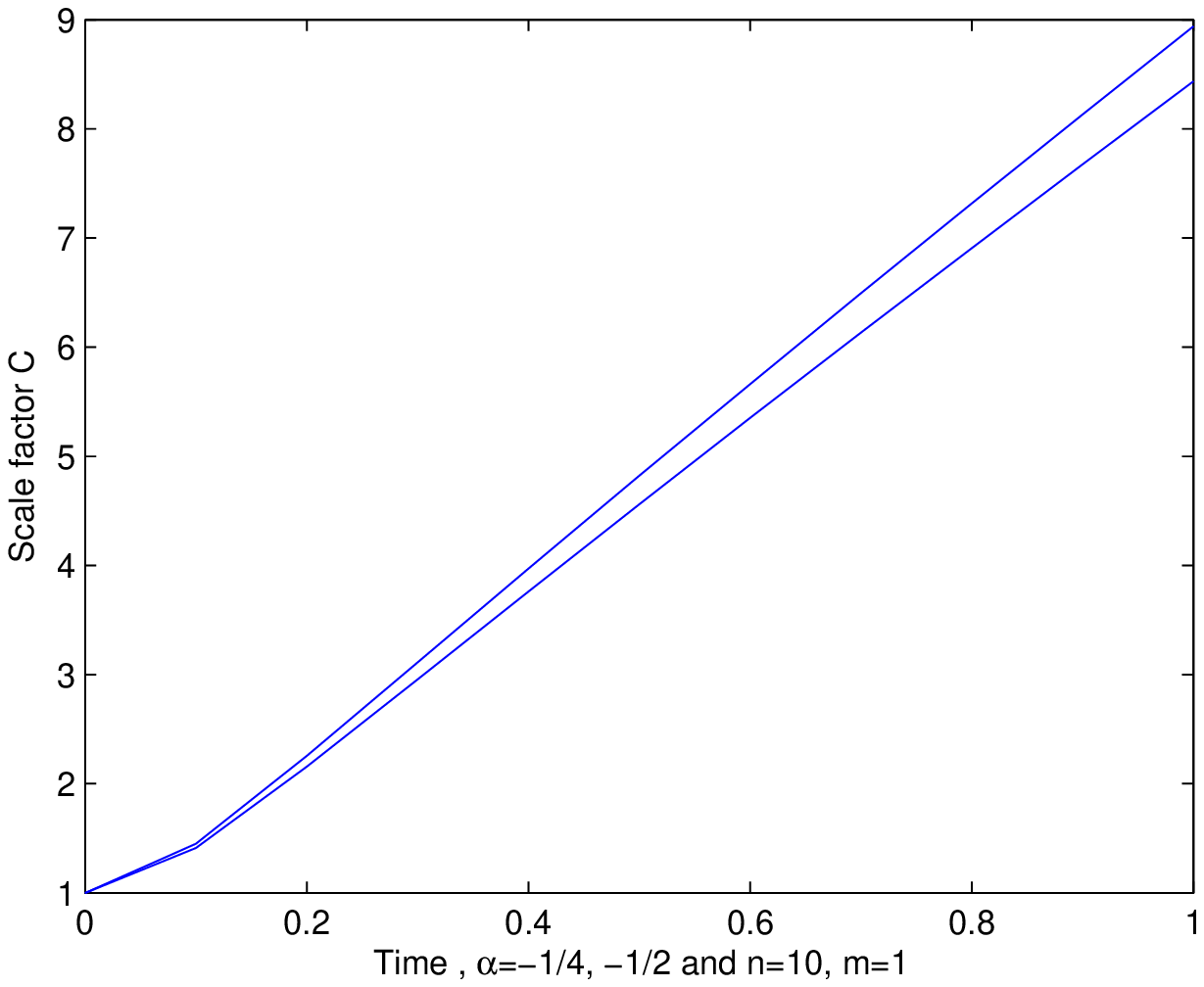}
\caption{Variation of the Scale factor $C$ with time $t$.}
\end{minipage}
\hspace{0.25cm}
\begin{minipage}{0.5\linewidth}
\includegraphics[width=0.95\textwidth]{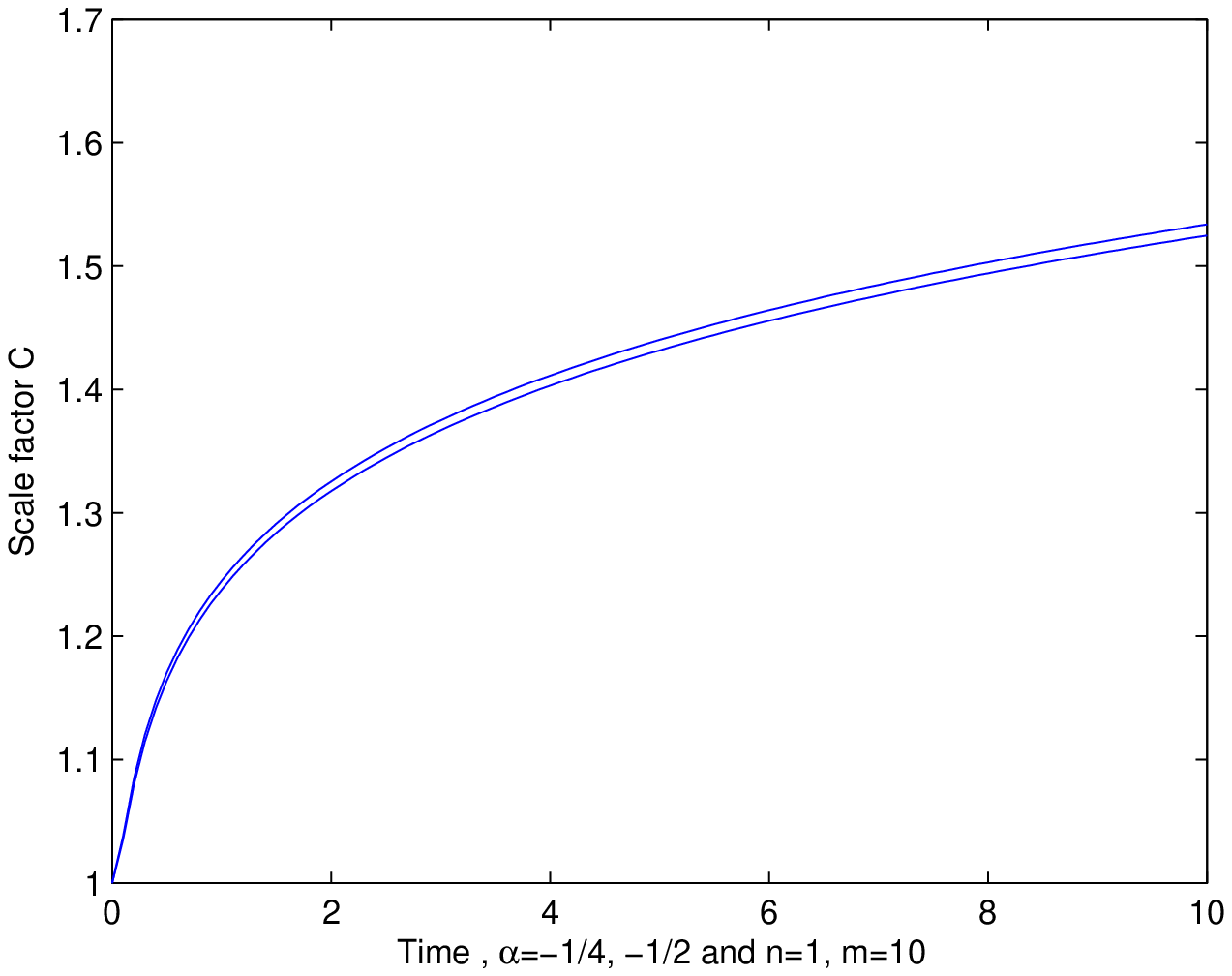}
\caption{Variation of the Scale factor $C$ with the time $t$.}
\end{minipage}
\label{fig:3}
\end{figure}
%
%

We now investigate the behavior of the above cosmological model by
analyzing the different physical parameters. The above set of exact
solutions shows that the expansion scalar $\theta$, shear scalar
$\sigma^2$ and the Hubble parameter $H$ are infinite at the time
$t=0$. At the same time $t=0$, all the directional Hubble's
parameters are also infinite. The pressure and density both will be
infinite at this epoch at $t=0$ iff $v_2 = 0$. These characteristics
of different physical parameters identify the existence of
singularity in the model at the initial time $t=0$. Now one can also
observe that all these parameters $\theta$, $\sigma^2$, $H$, $H_1$,
$H_2$, $H_3$, $\rho$ and $p$ are become zero at the large time
$t\rightarrow \infty$, even for $v_2 \ne 0$. That is all these
physical parameters are decreasing functions of time. Therefore this
model describes a continuously expanding and shearing universe with
the singularity at $t=0$.
This model gives an empty space for large time.\\

\noindent
Let us now study the behavior of the volume scalar and the scale factors
$A$, $B$, $C$ in this model. From Figure 1 , it can be seen that the volume
scalar $V(\alpha=-1/2, -1/4)$ is the increasing function of time. That is $V$
is zero at $t=0$ and it takes infinite value at $t\rightarrow \infty$. As $A$
is a function of $V$, namely $A = \sqrt{V}$ the behavior of $A$ is almost the
same as that of $V$. As far as $B$ and $C$ are concerned, depending on the
values of $m$ and $n$ they either expands rapidly or slowly. These variations
can be observed through the Figures 2, 3, 4, 5 and 6 respectively. \\

\noindent \textbf{4  Conclusion}\\

\noindent
In this paper, we have obtained an exact solution for the field equations of
Scale-Covariant theory of gravitation in Bianchi type VI line element of the
universe. Under some specific assumptions, exact solutions to the corresponding
field equations are found. It is found that one of the metric functions $(A)$
is an expanding one with acceleration whereas depending on the choice of the
parameters two other metric functions $B$ and $C$ expand either with acceleration
or deceleration. The model in question does not allow isotropization of the
initial anisotropic space-time. All the physical and kinematical parameters
have been thoroughly discussed. The solution so obtained, represents a
continuously expanding and shearing model of the universe with the singularity
at the initial time $t=0$. This model gives an empty space for large time.  \\

\noindent \textbf{5  Acknowledgments}\\

This work is partially supported by a joint Romanian-LIT, JINR,
Dubna Research Project 4163-6-12/13, theme no. 05-6-1060-2005/2013.

\vspace{0.8cm}
\noindent{\bf References}\\

\noindent Canuto, V. et al.: Phys.Rev.D.{\bf16}, 6(1977)\\
Canuto, V., Hsieh, S.H., Adams, P.J.: Phys.Rev.Lett. {\bf 39}, 8(1977)\\
Saha, B.: Phys.Rev.D.{\bf69}, 124006(2004)\\
Ram, S., Verma, M.K., Zeyauddin, M.: Chin.Phys.Lett. {\bf 26}, 089802(2009)\\
Reddy, D. R. K., Naidu, R. L., Adhav, K. S.: Astrophys. Space.Sci. {\bf 307}, 365(2007)\\
Beesham, A.: Class.Quantum.Grav.{\bf 3}, 481(1986)\\
Venkateswarlu, R., Kumar, P.K.: Astrophys. Space Sci. {\bf 298}, 403(2005)\\
Reddy, D.R.K., Patrudu, B.M., Venkateswarlu, R.: Astrophys. Space Sci. {\bf 204}, 155(1993)\\

\end{document}